\documentclass[%
twocolumn,
superscriptaddress,
nofootinbib,
 amsmath,amssymb,
 aps, prd
]{revtex4-2}

\usepackage{graphicx}
\usepackage{tensor}
\usepackage{siunitx}
\usepackage{xcolor} % Remove after notes are no longer needed
\usepackage[normalem]{ulem} % Remove after notes are no longer needed
\usepackage{lipsum} % Remove after filler text is no longer needed

\usepackage[pdfborder={0 0 0}, plainpages=false, hyperfigures=true]{hyperref}

\newcommand{\beq}{\begin{equation}}
\newcommand{\eeq}{\end{equation}}
\newcommand{\beqn}{\begin{eqnarray}}
\newcommand{\eeqn}{\end{eqnarray}}

\newcommand{\lo}{\mathrel{\raise.3ex\hbox{$<$}\mkern-14mu
    \lower0.6ex\hbox{$\sim$}}}
\newcommand{\go}{\mathrel{\raise.3ex\hbox{$>$}\mkern-14mu
    \lower0.6ex\hbox{$\sim$}}}

\newcommand{\UNH}{\affiliation{Department of Physics \& Astronomy, University of New Hampshire, 9 Library Way, Durham NH 03824, USA}}

\newcommand{\Wright}{\affiliation{ORCID: 0000-0002-5953-4221}}
\newcommand{\Fullerton}{\affiliation{Gravitational Wave Physics and Astronomy Center, California State University Fullerton, Fullerton, California 92834, USA}}
\newcommand{\UCBP}{\affiliation{Department of Physics, University of California, Berkeley, Berkeley, CA 94720, USA}}
\newcommand{\UCBA}{\affiliation{Department of Astronomy and Theoretical Astrophysics Center, University of California, Berkeley, Berkeley, CA 94720, USA}}
\newcommand{\LBL}{\affiliation{Nuclear Science Division, Lawrence Berkeley National Laboratory, Berkeley, CA 94720, USA}}
\newcommand{\NU}{\affiliation{Center for Interdisciplinary Exploration \& Research in Astrophysics (CIERA), Physics \& Astronomy, Northwestern University, Evanston, IL 60202, USA}}
\newcommand{\GRAPPA}{\affiliation{GRAPPA, Anton Pannekoek Institute for Astronomy and Institute of High-Energy Physics, University of Amsterdam, Science Park 904, 1098 XH Amsterdam, The Netherlands}}

\begin{document}
\title{Estimating outflow masses and velocities in merger simulations: impact of r-process heating and neutrino cooling}
\author{Francois Foucart}\UNH
\author{Philipp M\"osta}\GRAPPA
\author{Teresita Ramirez}\UNH\Fullerton\NU 
\author{Alex James Wright}\Wright
\author{Siva Darbha}\UCBP
\author{Daniel Kasen}\UCBP\UCBA\LBL

\begin{abstract}

The determination of the mass, composition, and geometry of matter outflows in black hole-neutron star and neutron star-neutron star binaries is crucial to current efforts to model kilonovae, and to understand the role of neutron star merger in r-process nucleosynthesis. In this manuscript, we review the simple criteria currently used in merger simulations to determine whether matter is unbound and what the asymptotic velocity of ejected material will be. We then show that properly accounting for both heating and cooling during r-process nucleosynthesis is important to accurately predict the mass and kinetic energy of the outflows.  These processes are also likely to be crucial to predict the fallback timescale of any bound ejecta. We derive a model for the asymptotic veloicity of unbound matter and binding energy of bound matter that accounts for both of these effects and that can easily be implemented in merger simulations. We show, however, that the detailed velocity distribution and geometry of the outflows can currently only be captured by full 3D fluid simulations of the outflows, as non-local effect ignored by the simple criteria used in merger simulations cannot be safely neglected when modeling these effects.
Finally, we propose the introduction of simple source terms in the fluid equations to approximately account for heating/cooling from r-process nucleosynthesis in future seconds-long 3D simulations of merger remnants, without the explicit inclusion of out-of-nuclear statistical equilibrium reactions in the simulations.

\end{abstract}

\maketitle

\section{Introduction}

Multimessenger observations of neutron star-neutron star (NSNS) and black hole-neutron star (BHNS) binaries provide us with information about the properties of dense matter~\cite{Flanagan2008,Read2009b,GW170817-NSRadius}, the origin of heavy elements~\cite{1976ApJ...210..549L,2017Natur.551...80K,2017Natur.551...67P}, the population of compact objects~\cite{Abbott:2020gyp}, their formation mechanism~\cite{Broekgaarden:2021iew}, and even the expansion rate of the Universe~\cite{Hotokezaka:2018dfi,Coughlin:2020ozl}. So far, one NSNS merger (GW170817) has been observed through both gravitational and electromagnetic waves~\cite{TheLIGOScientific:2017qsa,GBM:2017lvd,Monitor:2017mdv,2017Sci...358.1559K,2017ApJ...848L..19C,Cowperthwaite:2017dyu,2017Sci...358.1583K,2017ApJ...848L..32M,2017ApJ...848L..18N,2017Natur.551...75S,2017ApJ...848L..16S,2017ApJ...848L..27T,2017Sci...358.1565E,Mooley:2018dlz,Balasubramanian:2021kny}. A few additional mergers that likely involved at least one neutron star have been observed in gravitational waves only, most notably the likely NSNS merger GW190425~\cite{Abbott:2020uma} and the likely BHNS mergers GW200105 and GW200115~\cite{LIGOScientific:2021qlt}. A much larger number of likely binary mergers have also been observed for decades as short gamma-ray bursts~\cite{1993ApJ...413L.101K,1995SSRv...71..265A,Gehrels2009,2009ApJ...697.1071A,Fong:2015}, and a small number of these gamma-ray bursts were followed by optical/infrared transients (kilonovae) that may have been powered by radioactive decays in the matter ejected by BHNS and/or NSNS mergers~\cite{Tanvir:2013,Berger:2013,jin:15,Yang:2015pha}. 

While observing mergers through gravitational waves, gamma-ray bursts, or kilonovae alone is certainly valuable, the outsize scientific impact of GW170817 has clearly demonstrated the value of multimessenger observations of these systems. However, the interpretation of electromagnetic signals powered by NSNS and BHNS mergers remains difficult. Gamma-ray burst observations provide very limited information about the properties of the merging compact objects beyond the likely presence of a neutron star, in large part because the exact process powering these bursts remain poorly understood today. Kilonovae, which are powered by radioactive decay of the elements produced through r-process nucleosynthesis in neutron rich outflows~\cite{1976ApJ...210..549L,Li:1998bw,2010MNRAS.406.2650M,Roberts2011}, are in theory easier to connect to the masses, sizes, and spins of merging compact objects: the magnitude, color, and duration of a kilonova are strongly impacted by the mass, velocity, and composition of the matter ejected by a merger~\cite{2013ApJ...775...18B}; and the properties of these outflows can themselves be connected to the properties of the merging compact objects. Nevertheless, uncertainties in the numerical simulations used to model these outflows, in the properties of the elements produced during the r-process, and in the heating rate of the ejecta due to radioactive decays create significant difficulties when attempting to interpret kilonovae signals, with highly uncertain error bars in any inference made about the parameters of the merging compact objects (see e.g.~\cite{Barnes:2020nfi,Zhu:2020eyk,Raaijmakers:2021slr}). Material that is initially ejected by the merger yet remains bound to the post-merger remnant may also have a significant impact on late electromagnetic emission from the merger. In particular, late accretion onto the post-merger remnant may power x-ray emission in short gamma-ray bursts~\cite{2010MNRAS.402.2771M}, including in GW170817~\cite{2021arXiv210402070H,Metzger:2021grk}. The fallback timescale and associated x-ray lightcurves are likely to be impacted by heating/cooling of the ejecta during r-process nucleosynthesis~\cite{2010MNRAS.402.2771M,Desai:2018rbc,Ishizaki:2021qne}.

Predicting the properties of matter outflows in NSNS and BHNS mergers is crucial to our ability to perform detailed analysis of post-merger electromagnetic signals. Such predictions are typically made by developing analytical fits to the results of merger and post-merger simulations of these systems~\cite{Kawaguchi:2016,Dietrich:2016fpt,Coughlin:2018fis,Barbieri:2019sjc,Kruger:2020gig,Raaijmakers:2021slr}. A reliable determination of the properties of unbound material in merger simulations is accordingly important to current modeling efforts. 
Unfortunately, the finite size of computational grids, finite length of numerical simulations, and approximate treatment of nuclear physics in these simulations can make it difficult to robustly determine whether a given region of a simulation contains bound or unbound material. Historically, most general relativistic simulations have determined whether matter is bound or unbound by assuming that all matter beyond a certain radius follows geodesics of a time-independent metric~\cite{hotokezaka:13}. This ignores potentially important cooling (e.g. neutrino emission associated with $\beta$-decay of neutron-rich nuclei during r-process nucleosynthesis) and heating terms (e.g. thermalization of the energy released by the r-process), as well as the impact of pressure forces and time variations in the metric after the end of the numerical simulations. An alternative method, commonly used today, is to rely on the general relativistic Bernoulli criterion, which assumes a steady-state flow in a time-independent metric. This criterion partially accounts for the decompression of the fluid, and the conversion of internal and/or thermal energy into kinetic energy, but ignores neutrino cooling and the evolution of the metric in time. 

In practice, the Bernoulli criterion tends to overestimate the mass of unbound material while the geodesic criterion tends to underestimate it; and differences between the predictions of the two methods can be large. For example, the Bernoulli criterion can predict more than twice as much unbound matter than the geodesic criterion for hot outflows close to NSNS merger remnants~\cite{Kastaun:2014fna}. As the material expands and cools down, however, the two criteria will begin to agree. As the Bernoulli criteria accounts for the thermal energy of the outflows, one would expect it to perform better for hot outflows close to the remnant; the Bernoulli criteria has thus been particularly useful to study polar outflows in NSNS binaries (see e.g.~\cite{Vsevolod:2020pak}). For cold ejecta, on the other hand, it is not as clear which criterion performs best: some simulations have shown that the Bernoulli criterion flags as unbound material that in fact never expands enough to make use of its internal energy before falling back onto the remnant~\cite{Kawaguchi:2020vbf}.

As long as one neglects nuclear reactions in the outflows, these two criteria will both converge to the correct answer as matter moves away from the remnant. Accordingly, they do a good job of predicting matter outflows in simulations using simple, composition independent equations of state with limited microphysics. In a realistic merger, however, the energy released through r-process nucleosynthesis will be partially thermalized in the ejecta, and partially lost to neutrino emission~\cite{2010MNRAS.402.2771M}. The $\sim 3\,{\rm MeV}$ per nucleon deposited in the ejecta by r-process nucleosynthesis for neutron-rich matter ($Y_e\lesssim 0.1$) can have a significant impact on the dynamics of the outflows. It is for example enough to bring marginally bound matter to an asymptotic velocity $v\sim 0.08 c$. Including r-process heating thus leads to a correction to the asymptotic velocity of unbound matter~\cite{FoucartBhNs2016,Klion:2021jzr}, and to predictions for the mass of matter ejected by a merger. Whether r-process heating is sufficient to unbind marginally bound material is however a complex question that depends on how the orbital timescale and heating timescale compare~\cite{2010MNRAS.402.2771M}. 

Heating due to r-process nucleosynthesis is generally not directly included in general relativistic merger simulations: the timescale for the r-process is $\sim 1\,{\rm s}$, while merger simulations are typically continued for $\lesssim 0.1\,{\rm s}$ post-mergers. Only a few long-term simulations of the ejecta have been performed so far, including continuations of the smoothed particle hydrodynamics merger simulations of~\cite{Rosswog:2012fn,2013MNRAS.430.2585R} in~\cite{Rosswog2014,2014MNRAS.439..757G}, a post-processing of the general relativistic, grid-based BHNS merger simulations of~\cite{FoucartBhNs2016} in~\cite{Darbha:2021rqj}, and post-processing the evolution of the outflows observed in simulations of post-merger remnants~\cite{Fernandez:2018kax} in~\cite{Klion:2021jzr}. Using a careful implementation of the Bernoulli criterion, it is possible to partially correct for energy deposition from the r-process in merger simulations; e.g.~\cite{Kastaun:2014fna,Fujibayashi:2020dvr} use versions of the Bernoulli criteria that effectively thermalize $\sim 100\%$ of the energy released by the r-process in the ejecta. Accounting for both r-process heating and neutrino losses, however, requires more care in the definition of unbound matter. In this manuscript, we investigate a few potential methods to predict whether a given region of a simulation contains unbound material or not, including new criteria that attempt to account for neutrino cooling and thermalization efficiency in the outflows.

In Sec.~\ref{sec:methods}, we first review the geodesic criterion, as well as various implementations of the Bernoulli criterion that effectively account in different ways for r-process heating and thermalization. Based on this discussion, we then propose an improved version of the Bernoulli criterion that explicitly accounts for both r-process heating and thermalization efficiency, for any initial composition of the outflows. We note that we use a formalism that makes explicit its assumptions about the energy released by the r-process and the losses due to neutrino emission. This will allow us to modify the algorithm as needed when improved estimates for these quantities become available. We additionally rely on the study of r-process heating in initially bound outflows performed by Desai {\it et al}~\cite{Desai:2018rbc} to estimate what part of the initially bound matter is heated fast enough by the r-process to become unbound before reaching apoastron.  Finally, we discuss how our simple model can be adapted to approximately include both r-process heating and neutrino cooling in future seconds-long general relativistic simulations performed using standard finite volume methods, without the inclusion of complex out-of-nuclear statistical equilibrium nuclear reaction network.

In Sec.~\ref{sec:res}, we then compare the predictions of these different models for BHNS and NSNS mergers, and additionally compare these results to the outflows measured in a 3D smoothed-particle hydrodynamics simulation of BHNS merger outflows that does follow the outflows for multiple seconds and includes an explicit heating term~\cite{Darbha:2021rqj}. We argue that in relatively short merger simulations, including both future cooling and future heating terms is crucial to properly account for the energy budget of the outflows. On the other hand, we note that none of these simple models properly predict the detailed distribution of outflow velocities observed in hydrodynamics simulations, emphasizing the need to post-process simulations results to properly determine the geometry of merger outflows.

\section{Outflow models: unbound matter and asymptotic velocity}
\label{sec:methods}

\subsection{Geodesic criterion}

The simplest method often used in numerical simulations to estimate whether a fluid element
is unbound, and what its asymptotic velocity might be, is the geodesic criterion. When using that criterion, we make the assumption
that the ejecta follows spacetime geodesics in a time-independent, asymptotically
flat spacetime~\cite{hotokezaka:13}. Then, the time component of the 4-velocity one-form ($u_t$) is a conserved
quantity. A particle following a geodesic is unbound if $u_t<-1$, and its Lorentz factor at infinity 
($\Gamma_\infty$) is given by
\beq
\Gamma_\infty = -u_t.
\eeq
We note that if $\Gamma_\infty-1<0$, we can instead interpret that quantity as the specific gravitational
binding energy of the matter at late times, and use that estimate to obtain approximate values for the
orbital timescale of that bound material, and thus its fallback timescale. This will remain true for other estimates
of $\Gamma_\infty$ derived in this manuscript, and is discussed in more detail in Sections II.E and II.F.

A few milliseconds after merger, and when $GM/(rc^2)\ll 1$ (with $M$ the
mass of the remnant and $r$ the distance between the remnant and the ejecta), the assumption of a
time-independent, asymptotically flat spacetime is usually well justified. However, assuming that
particles follow spacetime geodesics is less accurate. Both thermal energy and nuclear binding energy
may be partially converted into kinetic energy as the fluid expands in the surrounding interstellar medium,
and that acceleration of the ejecta is entirely ignored by the `$u_t$' criterion. As a result, this simple method
typically seems to work well when (a) the ejecta is cold; and (b) simple equations of state that ignore nuclear binding energies
are used. 

We note that while the first condition may be physically correct for some outflows (most notably the dynamical ejecta
of BHNS mergers), the second always introduces an error in the inferred value of $\Gamma_\infty$.
The $u_t$ condition may appear to work well for cold ejecta in numerical simulations, in so far that $u_t$ appears
to remain constant as the ejecta moves away from the remnant, but this is somewhat misleading. Merger simulations
currently assume that the matter is in nuclear statistical equilibrium (NSE) at a given $Y_e$, while the r-process is an out-of-equilibrium process. Even if
merger simulations were able to follow the ejecta for the $\sim 1\,{\rm s}$ timescale over which r-process heating occurs,
they would simply predict that the outflows expand at constant composition (possibly up to minor composition changes due to neutrino-matter 
interactions) if using a composition-dependent equation of state, or at some predetermined 'equilibrium' composition for composition-independent equations of state (e.g. polytropes). Those simulations ignore any impact of an out-of-NSE evolution of the composition of the outflows,
and of the energy lost to neutrino emission during that out-of-NSE evolution.

\subsection{Bernoulli criterion version 1: constant composition}

A first improvement to the geodesics criterion is the Bernoulli criterion, used in many merger simulations today. The Bernoulli
criterion is often written as the assumption that a particle is unbound if $hu_t<-1$; and that, if it is unbound, its asymptotic Lorentz factor is
\beq
\Gamma_\infty = -hu_t.
\eeq
Here, $h=1+\epsilon+P/\rho$ is the enthalpy, $\epsilon$ is the specific internal energy, and $P$ is the pressure. In composition-dependent
equations of state, $\epsilon$ includes not only the thermal energy, but also the nuclear binding energy of the nuclei when in NSE at a given density, temperature, and composition. 

The Bernoulli criterion relies on the fact that $h u_t$ is constant along a given streamline of a steady-state flow. This is not formally applicable
to merger outflows, as these are not steady-state flows, yet the Bernoulli criterion appears to do a reasonable job of capturing the transformation 
of thermal energy and nuclear binding energy into kinetic energy during the expansion of the outflows (see e.g. Figs.1-3). As a result, it is probably the most broadly
used criterion in current simulations. It does however have important limitations.

The first issue is that, to use Bernoulli properly, we need to know the asymptotic enthalpy of the fluid $h_\infty$. In fact, in the form stated above, we implicitly 
assumed that $h_\infty=1$: if $hu_t$ is constant, the correct criteria for matter to be unbound is~\cite{Fujibayashi:2020dvr} 
\beq
h(-u_t) > h_\infty 
\eeq
and the asymptotic Lorentz factor is
\beq
\Gamma_\infty = -\frac{hu_t}{h_\infty}.
\eeq
Most of the simpler equations of state used in merger simulations (polytropes, piecewise polytropes, spectral) are composition independent, and satisfy $h_\infty=1$ by construction (asymptotically, $\rho\rightarrow 0$, $\epsilon\rightarrow 0$, $P/\rho\rightarrow 0$). For composition-dependent equations of state, on the other hand, the asymptotic enthalpy of the flow depends on the asymptotic composition of the flow, i.e.
\beq
\lim_{\rho\rightarrow 0} h = h_\infty(Y_{e,\infty})
\eeq
with $Y_{e,\infty}$ the asymptotic electron fraction (the temperature of the outflows is small at late times, and can be safely ignored in this expression). Typically, $h(\rho=0,T=0,Y_e)$ can vary by as much as $1\%$ with $Y_e$, due to the difference between the rest mass energy of free neutrons ($Y_e\sim 0$) and the rest mass energy per nucleon of the most strongly bound nuclei (e.g. $^{56}{\rm Ni}$). As merger simulations ignore out-of-equilibrium nuclear reactions, and thus have outflows with nearly constant $Y_e$ once $GM/(rc^2)\ll 1$, the criteria 
\beq
-h u_t>h(\rho=0,T=0,Y_e)
\eeq
with $Y_e$ the current electron fraction of the outflows in the simulation will perform very well at predicting the amount of matter that will be unbound by the merger within the limited set of physical processes included in the simulations. The formula
\beq
\Gamma_\infty = -\frac{hu_t}{h(\rho=0,T=0,Y_e)}
\eeq
thus provides good predictions for the asymptotic Lorentz factor under the same assumptions.

Unfortunately, physical outflows do not simply expand at constant $Y_e$. In reality, they undergo r-process nucleosynthesis, a process that (a) changes the electron fraction; (b) puts the outflows out of nuclear statistical equilibrium; and (c) leads to the loss of a significant amount of energy through emission of electron antineutrinos. Accordingly, no simulation explicitly uses this constant-$Y_e$ model to determine $\Gamma_\infty$. The energy released by the r-process is $\sim 7\,{\rm MeV}$ per nucleons (for neutron rich outflows), and about half of that energy goes into the escaping neutrinos. Both r-process heating and neutrino cooling are thus worth taking into account: an energy difference of $7\,{\rm MeV}$ per nucleon can be the difference between matter being marginally bound, and an ejecta with asymptotic velocity $v\sim 0.13c$!

\subsection{Bernoulli criterion version 2: r-process heating}
\label{sec:berheat}

To modify the constant-$Y_e$ Bernoulli criterion, let us first attempt to tackle r-process heating, ignoring losses due to neutrino emission during the r-process. This is reasonably simple to do, if we know the average binding energy of the nuclei created by the r-process. Indeed, asymptotically $P/\rho=0$, and thus if we know the asymptotic $\epsilon_\infty$, we know $h_\infty$ and can use
\beq
\Gamma_\infty = -\frac{hu_t}{h_\infty}
\eeq
As $h_\infty$ does not strongly depend on the 
initial properties of the ejecta, this criteria is actually simpler than Bernoulli without heating: we only need to know the current values of $h$ and $u_t$ in the ejecta, and the value of $h_\infty$. This is a heating term because, for neutron rich matter, $h(\rho=0,T=0,Y_e)>h_\infty$. This is practically very close to the criterion proposed by Fujibayashi {\it et al}~\cite{Fujibayashi:2020dvr}\footnote{In that work, $h_\infty$ is replaced by the minimum value of the function $h(\rho=0,T=0,Y_e)$, which differs from $h_\infty$ at the level of the difference between the binding energy of the ashes of the r-process and the binding energy of the most bound nuclei.}, as well as to the effective meaning of the condition $\Gamma_\infty = -hu_t$ in composition-independent equations of state\footnote{As composition-independent equations of state assume NSE and neutrinoless beta-equilibrium, leading to a value of $h(\rho=0,T=0)$ close to the minimum of  $h(\rho=0,T=0,Y_e)$, and thus tp results similar to Fujibayashi {\it et al}'s method~\cite{Fujibayashi:2020dvr}.}

We note that, in simulations, there is a subtlety in the calculation of $h_\infty=1+\epsilon_\infty$. The specific internal energy $\epsilon_\infty$ is just the average binding energy of the nuclei formed by the r-process; but the binding energy has to be defined with respect to a reference value $m_*$ for the mass of a nucleon. Different equations of state commonly used in numerical simulations make different choices for $m_*$. For example, the DD2 and SFHo equations of state from~\cite{2013ApJ...774...17S} use $m_* = 1u$, with $u$ the atomic mass unit; but the LS220 equation of state~\cite{Lattimer:1991nc} uses a mass closer to the mass of a neutron. The result of simulations is clearly independent of that choice; $m_*$ only provides us with an arbitrary separation between what we call the `baryon rest mass energy density' of the fluid ($\rho=nm_* c^2$, with $n$ the baryon number density) and the internal energy density of the fluid ($u=\rho \epsilon$). The baryon number (in general) and total energy (in time-independent spacetimes) have to satisfy conservation laws, but the separation between baryon mass and internal binding energy is somewhat arbitrary. What is sometimes called the `rest mass conservation' equation in relativistic simulations is in fact the equation for baryon number conservation, rescaled by $m_*$. 

As a result, with the DD2 or SFHo equation of state we have $h_\infty\sim 1$, but with the LS220 equation of state we have $h_\infty\sim 0.992$. Using $\Gamma_\infty = -hu_t$ with the LS220 equation of state leads to an underestimate of the unbound mass and of the velocity of the ejecta, unless the reference mass has been appropriately modified. We note that an error of $\sim 0.008$ in $h_\infty$ results in an error of $\sim 7\,{\rm MeV}$ per nucleon in the final kinetic energy of the ejecta, a non-negligible difference.

\subsection{Bernoulli criterion version 3: r-process heating and neutrino losses}

The last important effect that we would like to take into account is energy losses due to neutrino emission during r-process nucleosynthesis. We note that during the first few seconds of evolution, when most of the r-process energy is released, we can safely assume that only neutrinos escape the ejecta, while other products of nucleosynthesis are thermalized. If a fraction $f_{\rm loss}$ of the fluid's rest mass energy is lost to neutrinos during the r-process, we can use the condition
\beq
-hu_t (1-f_{\rm loss})>h_\infty
\label{eq:hC}
\eeq
to flag unbound material, and 
\beq
\Gamma_\infty=-\frac{hu_t}{h_\infty} (1-f_{\rm loss})
\eeq
for the asymptotic Lorentz factor. The value of $f_{\rm loss}$ is uncertain. Metzger {\it et al}~\cite{2010MNRAS.402.2771M} first posited that about $50\%$ of the energy released by the r-process at early times is thermalized in the ejecta. More recently Desai {\it et al}~\cite{Desai:2018rbc}, using the output of the nuclear reaction code SkyNet for BHNS merger outflows~\cite{Lippuner2015,roberts:17} and assuming that $45\%$ of the energy released by the r-process is lost to neutrinos, found a quasi linear relation between $Y_e$ and the heating rate (for $Y_e<0.2$) which, in our notation, would be 
\beq
f_{\rm loss} \approx 0.0032 - 0.0085 Y_e.
\eeq
Overall, we thus get the corrected expression
\beq
\Gamma_\infty=-\frac{hu_t}{h_\infty} \left(0.9968+0.0085 Y_e\right).
\label{eq:Ginf}
\eeq
A potential issue with this formula is that $f_{\rm loss}$ goes to zero for $Y_e=0.376$, and becomes negative when $Y_e>0.376$. For $Y_e>0.376$, we instead set $f_{\rm loss}=0$. Physically, we do not expect a true r-process for $Y_e \gtrsim 0.4$ (see e.g.~\cite{1992ApJ...395..202W}); at larger electron fractions, seed nuclei in the ejecta have an electron fraction comparable to the average electron fraction of the fluid, and there is thus no true neutron excess of the fluid  compared to those seed nuclei. While out-of-NSE reactions still occur within the ejecta, the difference in binding energy and $Y_e$ between the nuclei in the fluid in NSE at the original $Y_e$ and the nuclei actually produced at the end of the true out-of-NSE evolution of the system is significantly smaller than for more neutron-rich outflows, and neglected at the level of accuracy of our simple model.

We also note that this expression for $\Gamma_\infty$ still neglects non-local effects, but should at least perform better when it comes to calculating the total kinetic energy of the outflows than the expressions obtained in the previous sections. 

\subsection{Unbound criterion: Impact of the heating timescale}

Whether the expression derived in the previous section does well at predicting the mass of the outflows is uncertain, in part because we assume that there is enough time to heat marginally bound material before it starts falling back to the post-merger remnant, interrupting the r-process~\cite{2010MNRAS.402.2771M}. Desai {\it et al}~\cite{Desai:2018rbc} find that there is enough time to heat the ejecta for massive remnants ($M\gtrsim 12M_\odot$, with some dependence on the heating rate). For lower mass systems, it is possible to estimate which material will receive enough r-process energy to be unbound if we assume that the heating is roughly constant in time and distributed over a time $t_{\rm heat}$. If the initial binding energy per nucleon of the fluid is $E_0$ and its binding energy per nucleon at apoastron (after heating, for bound matter) is $E_f$, we should have
\beq
E_0 - E_f = Q \frac{t_{\rm orb}(E_f)}{2t_{\rm heat}}
\label{eq:dEvQ}
\eeq
with $Q$ the total r-process heating per nucleon and $t_{\rm orb}(E)$ the orbital timescale of fluid with binding energy $E$. The factor of $2$ in this expression is included because the heating needs to happen before the ejecta reaches apoastron. After apoastron, the density of the ejecta will start increasing, at which point the r-process is expected to end~\cite{2010MNRAS.402.2771M}. This expression is of course only valid if $t_{\rm orb}(E_f)<2t_{\rm heat}$. If not, we should set $E_f=E_0-Q$, as the r-process can be fully completed. The orbital timescale is~\cite{Desai:2018rbc}
\beq
t_{\rm orb}(E) = 1.6\,{\rm s} \left(\frac{E}{1\,{\rm MeV}}\right)^{-1.5} \left(\frac{M_{\rm rem}}{5M_\odot}\right),
\label{eq:tOrb}
\eeq
with $M_{\rm rem}$ the gravitational mass of the post-merger remnant. Our equation for $E_f$ only has a solution for
\beq
E_0 > E_{0,\rm lim} = 1.96 \left(\frac{Q}{1\,{\rm MeV}} \frac{0.8\,{\rm s}}{t_{\rm heat}} \frac{M_{\rm rem}}{5M_\odot}\right)^{0.4}\,{\rm MeV}.
\eeq
Material with a lower initial binding energy receives the full r-process heating $Q$. The condition for material to be unbound can then be approximated as
\beq
\frac{-hu_t}{h(\rho=0,T=0,Y_e)} > 1 -\frac{\rm{\min}(E_{0,\rm lim},Q)}{m_p c^2}
\label{eq:unb}
\eeq
with $m_p$ the mass of a proton. Note the use of $h(\rho=0,T=0,Y_e)$ and not $h_\infty$ here, as heating/cooling during the r-process is accounted for in the right hand side. For our choice of $f_{\rm loss}$ above, i.e. assuming $45\%$ of the r-process energy is lost to neutrinos, and using the same heating rate estimates as in~\cite{Desai:2018rbc}, we get
\beq
Q = (3.669-9.745 Y_e)\,{\rm MeV}
\eeq
for $Y_e<0.2$\footnote{While we include 3 significant digits in this formula, we note that the uncertainty in $Q$ is much larger than $10^{-3}$.}. A reasonable choice for the heating timescale is $t_{\rm heat}\approx 1\,{\rm s}$. We again assume that this expression is correct as long as it predicts $Q>0$, and we set $Q=0$ otherwise. As long as we have a good estimate of $M_{\rm rem}$, these expressions are relatively easy to use in numerical simulations.

We now have at our disposal a model that approximately includes r-process heating, neutrino cooling, thermal expansion, but still neglects non local effects. The asymptotic Lorentz factor in this model is given by Eq.~\ref{eq:Ginf}, and the condition for matter to be unbound by Eq.~\ref{eq:unb}. These equations rely on the choice of a uniform r-process heating timescale $t_{\rm heat}$ and a total r-process heating $Q(Y_e)$ (excluding the energy lost to neutrino emission) which are easy to modify when our understanding of their correct physical values improves. 

\subsection{Fallback timescale for bound material}

Very similar methods can be used to estimate the fallback timescale of material that remains bound to the remnant.  For the models that do not account for
the timescale required for r-process heating to occur, we simple assume $E=(1-\Gamma_{\infty})$, and obtain the fallback timescale
\beq
t_{\rm fallback} = t_{\rm orb}(E)
\eeq
using Eq.~\ref{eq:tOrb}. For the models that does account for the finite time required for r-process heating to happen, we instead need to jointly solve equations~\ref{eq:dEvQ}-\ref{eq:tOrb} for $E_f$ and $t_{\rm orb}(E_f)$, and interpret $t_{\rm orb}(E_f)$ as the fallback timescale of the material. This is
possible for any material with $E_0>E_{0,\rm lim}$. We note that if the solution $E_f$ satisfies $E_f<E_0-Q$, then we have to set $E_f=E_0-Q$ (the total heating per nucleon cannot exceed $Q$). This might happen if $t_{\rm orb}(E_f)>2t_{\rm heat}$, i.e. if the r-process timescale is shorter than the time required for a fluid element to reach apoastron. We also set $E_f=E_0-Q$ if  $E_0<E_{0,\rm lim}$, as the absence of a solution to equations~\ref{eq:dEvQ}-\ref{eq:tOrb} indicates that the r-process continues to completion.

\subsection{Toy model for numerical simulations}

Going further than these estimates will require numerical simulations of the outflows. Long-term simulations (a few seconds long) following ejected material while accounting for r-process heating may serve two important purposes. First, they would allow for more robust descriptions of the geometry of the outflows powering kilonovae~\cite{Darbha:2021rqj}, and of the velocity of these outflows. Second, for marginally bound ejecta, they would allow for more accurate predictions of the fallback timescale of the ejected matter, as well as for studies of potential deviations of the accretion rate from a simple power-law decay that could have important consequences for e.g. late-time x-ray emission in neutron star mergers~\cite{2010MNRAS.402.2771M,Desai:2018rbc}. 

It is in fact possible to emulate the heating due to the r-process in an approximate yet straightforward manner in simulations that evolve the general relativistic equations of fluid dynamics in conservative form. To do so, we note that our formulae predict that $Q=0$ for $Y_e\approx 0.38$. We also note that, looking at the output of the nuclear reaction network Skynet for r-process nucleosynthesis in neutron rich ejecta~\cite{Lippuner2015,roberts:17}, the final electron fraction of the ashes of the r-process is $Y_{e,f}\approx 0.38$. We will thus assume that, for $Y_e<0.38$, the r-process drives $Y_e$ to that final value on a timescale $t_{rp}$. This can be done by the addition of a source term to the equation of lepton number conservation
\beq
\frac{d(\rho_* Y_e)}{dt} = ... - \rho_* \frac{Y_e-Y_{e,f}}{t_{rp}}
\eeq
with $\rho_*=\rho \sqrt{-g} u^t$, $u^t$ the time component of the fluid 4-velocity, and $g$ the determinant of the spacetime metric.
As the electron fraction evolves in time, the r-process releases energy into the fluid. This is already partially taken into account by the equation of state, when using a nuclear equation of state: the equation of state assumes nuclear statistical equilibrium, and to first order the binding energy of the ashes of the r-process is comparable to the binding energy of the nuclei formed in NSE (the difference is typically $\sim 10\%$ of the energy released by the r-process). What is not yet taken into account is the energy lost to neutrinos emitted during the r-process. Accordingly, we need to include a {\it cooling term} in the evolution equations of the fluid
\beq
\nabla_\mu T^{\mu\nu} = -\dot Q_r u^\nu
\eeq
and not a heating term. Consistency with our earlier choices for $Q, f_{\rm loss}, t_{\rm heat}$ would lead to the formula
\beq
\dot Q_r = 0.0085 \rho_0 \frac{Y_e-Y_{e,f}}{t_{rp}}.
\eeq
and $t_{rp}\sim (0.5-1)\,{\rm s}$. If included in merger simulations continued for a few seconds post merger, these additional source terms should provide a reasonably good approximation to the impact of nuclear heating on the dynamics of the outflows, including non-local effects -- at least up to the $\sim 10\%$ difference between the binding energy of the ashes of the r-process and the binding energy of nuclei in NSE, as well as up to the impact of a more complex time-dependence of the r-process heating.

Adding these simple source terms to numerical simulations evolved for a few seconds post-merger could thus significantly improve the accuracy of the predicted outflow properties. We have verified using single-cell simulations that these simple evolution equations provide heating comparable to the output of the SkyNet nuclear reaction network for initially neutron rich ejecta. We note however that this model does not account for heating on timescales $\gg 1\,{\rm s}$. Heating on longer timescales does not significantly impact the dynamics of the outflows, but it is the main source of energy powering kilonovae. At late times, the heating is expected to be more weakly dependent on the initial $Y_e$, but on the other hand the thermalization efficiency of particles produced through nuclear reactions is more complex to model~\cite{Barnes:2016}. Our model could thus be used to get outflows to their nearly homogeneous expansion phase, but more advanced heating terms and thermalization models will still be needed when studying the production of kilonovae and when calculating lightcurves. 

\section{Impact on measured simulation outflows}
\label{sec:res}

To illustrate the impact of these different assumptions, we now post-process the outflows produced in a NSNS merger, in the early post-merger remnant of a NSNS binary, and in a BHNS merger, using different prediction methods. We also compare the BHNS results with the final velocity observed in Darbha {\it et al}~\cite{Darbha:2021rqj}, where one of our BHNS merger simulations is continued using the smoothed particle hydrodynamics code PHANTOM~\cite{Price:2017mwk}, and an approximate model for r-process heating is included in the simulation. 

The models considered here are:
\begin{itemize}
\item Model A:  $\Gamma_\infty = -u_t$, with unbound criterion $u_t<-1$ (geodesic criterion)
\item Model B: $\Gamma_\infty = -h u_t/h_\infty$, with unbound criterion $hu_t<-h_\infty$ (Bernoulli without neutrino losses)
\item Model C: $\Gamma_\infty=-\frac{hu_t}{h_\infty} \left(0.9968+0.0085 Y_e\right)$, with unbound criterion $\Gamma_\infty>1$ (Bernoulli with neutrino losses)
\item Model D: $\Gamma_\infty=-\frac{hu_t}{h_\infty} \left(0.9968+0.0085 Y_e\right)$, with unbound criterion
\beq
-hu_t>h(\rho=0,T=0,Y_e)\left(1 -\frac{\rm{min}(Q,E_{0,\rm lim})}{m_p c^2}\right).
\eeq
(Bernoulli with neutrino losses and with corrections for the finite duration of an orbit)
\end{itemize}
For the BHNS merger, we also consider an additional Model E that includes the same amount of heating as the PHANTOM simulation. As the PHANTOM simulation only evolves fluid elements with $u_t<-1$ in the merger simulation, we use Model E with the unbound criteria $u_t<-1$, for consistency.

In this manuscript, we mostly limit ourselves to the study of unbound material, although we provide information about the fallback timescale of bound material
for models A-D in our BHNS merger simulation. The impact of r-process heating and neutrino cooling on bound matter is very dependent on the parameter of the system, and may lead to an early shut-down of fallback accretion or time gaps in the accretion history of the post-merger remnant. We refer the reader to~\cite{Desai:2018rbc} for a more detailed discussion of this process across the range of parameters expected in BHNS and NSNS mergers.

\subsection{Neutron star-neutron star merger}

\begin{figure}
\includegraphics[width=0.99\columnwidth]{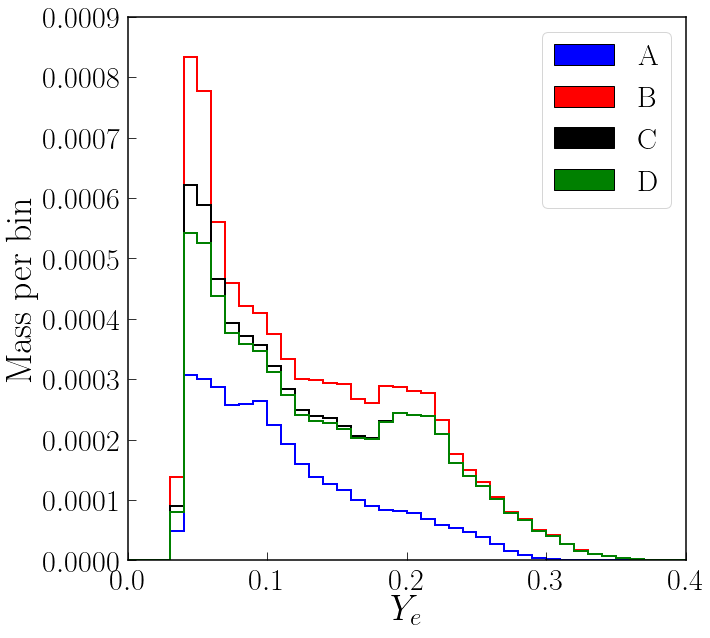}\\
\includegraphics[width=0.99\columnwidth]{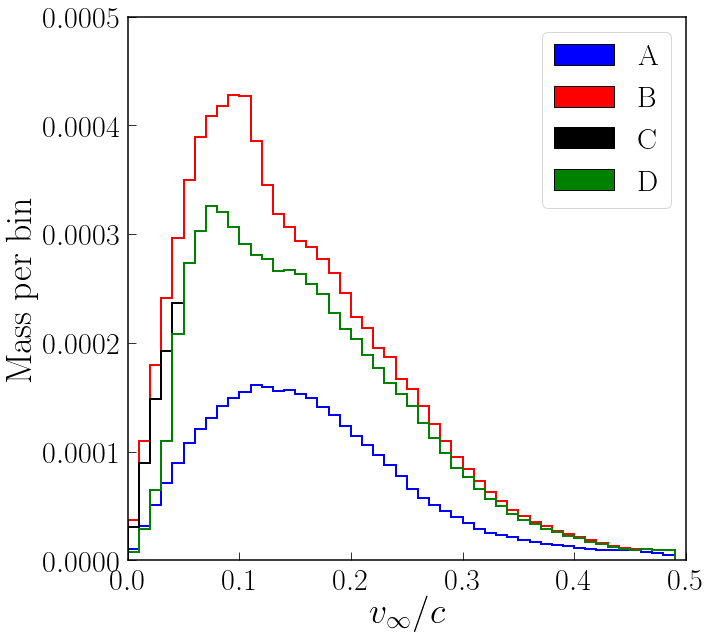}
  \caption{Predicted mass of unbound matter as a function of $Y_e$ ({\it Top}) and of the asymptotic speed $v_\infty$ ({\it Bottom}) in a NSNS simulation for our four outflow models.}
\label{fig:NsNs}
\end{figure}

We begin with the outflows observed in a NSNS merger simulation~\cite{Foucart:2020qjb} performed with the SpEC merger code~\cite{SpECwebsite}. That simulation includes Monte-Carlo radiation transport, but not magnetic fields. It uses neutron stars of mass $m_1=1.58M_\odot$, $m_2=1.27M_\odot$, and the DD2 equation of state. The simulation ends $5\,{\rm ms}$ after merger. The total mass of material flagged as unbound is $M_{\rm ej}=(0.0034,0.0083,0.0068,0.0066)M_\odot$ for models $(A,B,C,D)$ respectively, while the estimated asymptotic kinetic energy of the ejecta is $K_{\rm ej}=(6.6,14.1,12.5,12.5)\times 10^{-5}M_\odot c^2$. Model $A$, which ignores thermal energy and nuclear binding energy in the outflows, vastly underestimates the ejected mass. Model B, which ignores cooling, overestimates the ejected mass. The difference between models C and D is smaller, which is probably for the best considering that the unbound criteria used in model D is the most ad-hoc part of our estimates. From these results, it would be reasonable to assume that $\sim (0.0065-0.0070)M_\odot$ is ejected in our simulation (ignoring finite resolution errors for now...). 

Fig.~\ref{fig:NsNs} provides more information about the initial electron fraction and asymptotic velocity of the outflows according to these four models. Model A underestimates the outflows across the entire parameter space, but model B,C,D are in good agreement for less neutron-rich ejecta, when nuclear heating plays a smaller role in the dynamics of the outflows. Models C and D mostly disagree for low-$Y_e$, low-velocity outflows: these outflows have the largest amount of r-process heating, and may end up bound or unbound depending on when / how fast the r-process energy is deposited in the fluid.

\subsection{Neutron star-neutron star merger remnant}

\begin{figure}
\includegraphics[width=0.99\columnwidth]{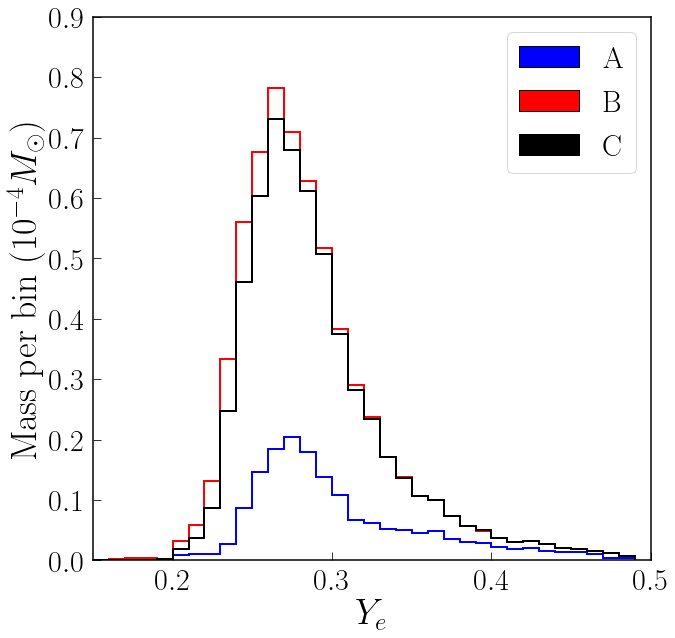}\\
\includegraphics[width=0.99\columnwidth]{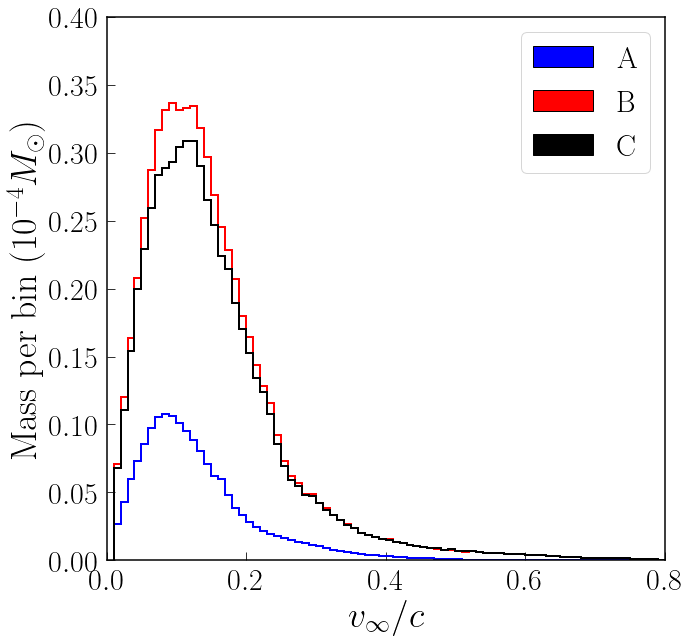}
  \caption{Predicted mass of unbound matter as a function of $Y_e$ ({\it Top}) and of the asymptotic speed $v_\infty$ ({\it Bottom}) for the remnant of a  NSNS merger $32\,{\rm ms}$ post-merger. We show all outflow models except model D, which is expected to behave very similarly to model C.}
\label{fig:Remnant}
\end{figure}

Next we analyze the outflows observed in a NSNS merger
remnant simulation presented in \cite{moesta:20} and performed
with the open-source \texttt{Einstein Toolkit}~\cite{Loffler:2011ay} module
\texttt{GRHydro}~\cite{moesta:14a}. The simulation included a magnetic field
and employed the $K_0 = 220\,\mathrm{MeV}$ variant of the equation of state
of \cite{Lattimer:1991nc} and the neutrino leakage/heating approximations described
in \cite{OConnor2010} and \cite{ott:13}. The simulation was carried out until the
remnant hypermassive neutron star collapses to a black hole (approximately 42ms after merger).
The data analyzed in this manuscript considers all matter flagged as unbound by a model $32\,{\rm ms}$ after merger.
We only consider models A,B,C. As in other examples, we expect that the results of model D would be very similar to those of model C. 

Models A,B and C predict an unbound mass on the grid of, respectively, $M_{\rm ej}=(1.7,6.3,5.8)\times 10^{-4}M_\odot$; they also predict total asymptotic kinetic energy $K_{\rm ej}=(2.1,11.9,11.5)\times 10^{-6}M_\odot c^2$. We see that, in this post-merger remnant, model A still vastly underestimates the amount of matter unbound by the system. This is not a surprising results: post-merger outflows typically have higher enthalpy than the cold, neutron-rich dynamical ejecta; model A is thus very inaccurate when used on these outflows. On the other hand, models B and C are in much better agreement. We can improve our understanding of the reasons behind this agreement by looking at Fig.~\ref{fig:Remnant}. Most of the ejecta has $Y_e\gtrsim 0.25$, and thus releases a lot less energy during the early stages of r-process nucleosynthesis than the very neutron-rich tidal ejecta studied in the previous section. 

\subsection{Black hole-neutron star binary}

\begin{figure}
\includegraphics[width=0.99\columnwidth]{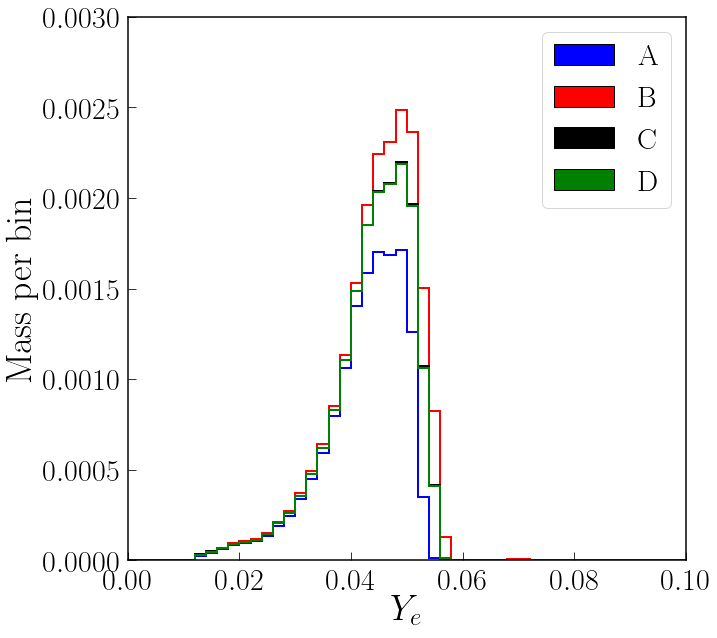}\\
\includegraphics[width=0.99\columnwidth]{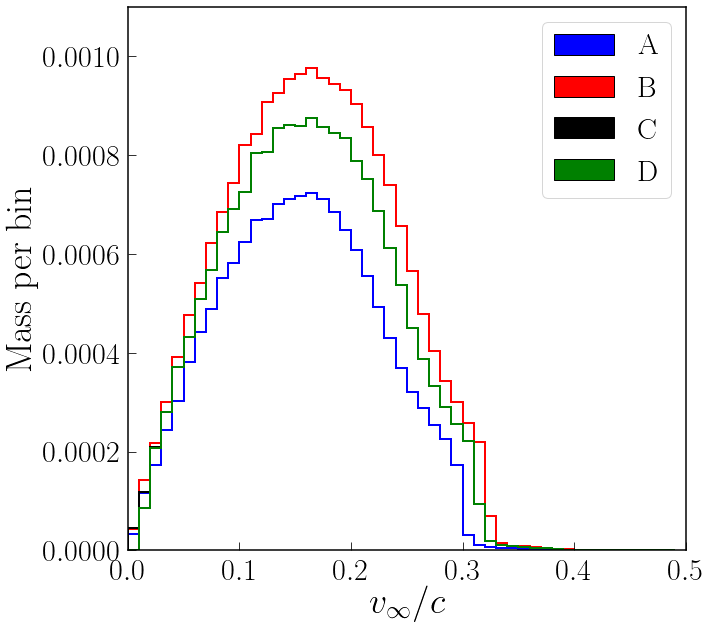}
  \caption{Same as Fig.~\ref{fig:NsNs}, but for a BHNS binary.}
\label{fig:BhNs}
\end{figure}

Finally, we consider a BHNS binary merger. We use simulation M14-M5-S9-I60 from~\cite{FoucartBhNs2016}, the merger of a $1.4M_\odot$ neutron star and a $5M_\odot$ black hole with dimensionless spin $\chi_{\rm BH}=0.9$ initially inclined by $60^\circ$ with respect to the orbital angular momentum of the binary. The properties of the outflows are measured $5\,{\rm ms}$ after merger. In this simulation, neutrinos are modeled using a simple leakage scheme, while magnetic fields are ignored. For this system, we predict an ejected mass $M_{\rm ej}=(0.0140,0.0200, 0.0176,0.0175)M_\odot$ for models $(A,B,C,D)$. There is better agreement on the ejected mass than in the NSNS binary because there is nearly no marginally bound ejecta, and the ejecta is relatively cold. We find larger variations in the estimated kinetic energies of the ejecta, $K_{\rm ej}=(20.7,33.9,28.4,28.4)\times 10^{-5}M_\odot c^2$. Again, models C and D are in very good agreement. The predicted $Y_e$ and $v_\infty$ distributions of the ejecta for each model are shown in Fig.~\ref{fig:BhNs}. The $Y_e$ distribution is always narrow and centered on $Y_e\sim 0.05$. The main impact of including r-process heating/cooling is to shift the tail of the velocity distribution, and to increase the mass of the outflows. Models C and D are indistinguishable on the figure.

\begin{figure}
\includegraphics[width=0.99\columnwidth]{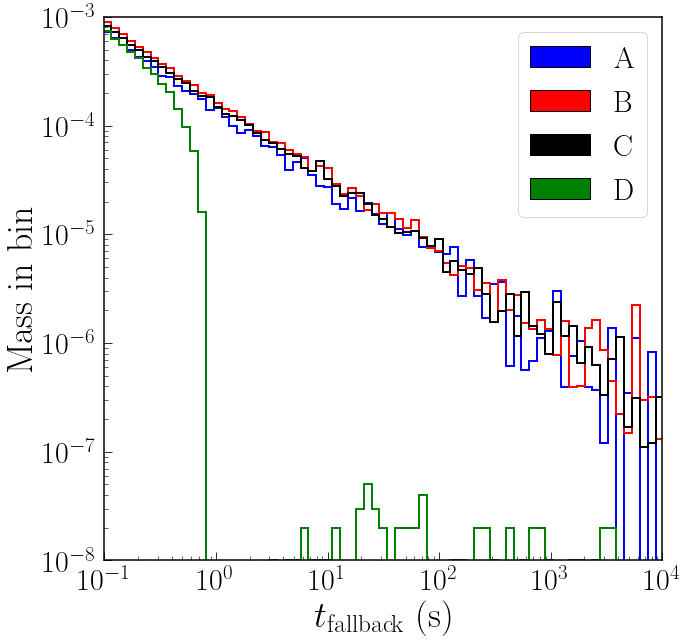}\\
  \caption{Comparison of predictions for the distribution of fallback times in a BHNS binary merger, using the same models as in Fig.~\ref{fig:BhNs}.}
\label{fig:BhNsFallback}
\end{figure}

For this BHNS merger, we also investigate the impact of different models on the predicted fallback time of bound material onto the remnant. The results are
shown in Fig.~\ref{fig:BhNsFallback}. We see that in this case, model D is significantly different from models A,B, and C. We can understand this rather easily by 
noting that the main differences between models A,B, and C result in a smooth shift of the distribution function of the ejecta as a function of $\Gamma_\infty$. As that distribution function does not vary rapidly around $\Gamma_\infty=1$, the amount of marginally bound material (i.e. material falling back on timescales $\gtrsim 1\,{\rm s}$) is relatively insensitive to the choice of model. Model D, on the other hand, creates a gap in that distribution function between material with a truncated r-process (material falling back onto the remnant within $\lesssim 1\,{\rm s}$ after the merger), and material that undergoes the full r-process, which mostly ends up unbound for the parameters chosen here (a very small amount of matter remains marginally bound). This cut-off in the accretion rate of the post-merger remnant is discussed in more detail in~\cite{Desai:2018rbc}, together with other potential consequences of a truncated r-process (e.g. temporal gaps in the accretion rate).

\begin{figure}
\includegraphics[width=0.99\columnwidth]{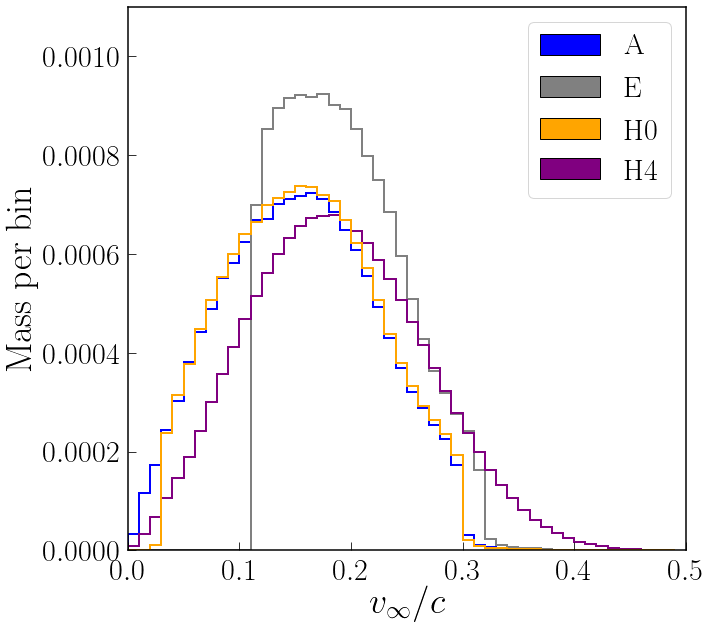}\\
  \caption{Comparison of predictions for the asymptotic velocity of the ejecta between simple models post-processing merger simulations without ($A$) and with ($E$) heating, and the result of a hydrodynamics simulation with heating included ($H4$), and without heating ($H0$).}
\label{fig:BhNsPhantom}
\end{figure}

Finally, we compare our results for unbound matter to PHANTOM simulations of the same system~\cite{Darbha:2021rqj}. The PHANTOM simulations are initialized from the snapshot of the SpEC merger simulation analyzed in this section, but only regions with $u_t<-1$ are evolved. In~\cite{Darbha:2021rqj}, multiple heating models are considered; here we compare results to the H0 model (no heating) and the H4 model (r-process heating of $5.9\,{\rm MeV}$ per nucleon, i.e. order-of-magnitude accurate but a bit larger than the most likely heating rates). We compare these two results to model A (no heating) and model E ($\Gamma_\infty = -u_t (1+5.9\,{\rm MeV}/(m_p c^2))$). All models agree on the mass of the ejecta, by construction (they all use the $u_t<-1$ criteria). The predicted kinetic energy of the ejecta is $K_{\rm ej}=(21.3,29.8)\times 10^{-5}M_\odot c^2$ for models (H0,H4) of the PHANTOM simulations and $K_{\rm ej}=(20.0,29.6)\times 10^{-5}M_\odot c^2$ for the post-processing models $(A,E)$. We see that given the same total heating, the full hydrodynamics simulation and the post-processing agree on the total kinetic energy of the ejecta. However, non-local effects do impact the distribution of velocities. Fig.~\ref{fig:BhNsPhantom} shows $v_\infty$ for models $(A,E,H0,H4)$. We see that while the average velocities of models $E$ and $H4$ are comparable, hydrodynamical effects cause the production of a much broader velocity distribution in the $H4$ model. The outflows also become more spherical, although this only becomes obvious when looking at their spatial distribution. These changes were already noted in~\cite{Darbha:2021rqj}, and similar effects on the geometry of the outflows in the presence of r-process heating were observed in~\cite{Rosswog2014}. The sharp cutoff in velocity at $v_\infty\sim 0.1c$ for $E$ is an artifact of the unbound condition $u_t<-1$, taken to match the assumptions of the H4 simulation rather than for its realism. The simulation without heating ($H0$) nearly exactly agrees with the predictions of the model without heating ($A$), indicating that the spread in the velocity observed in model $H4$ is truly due to non-local redistribution of the energy released by heating, rather than simply to the use of a hydrodynamical simulation. 

\section{Conclusions}

In this manuscript, we reviewed the various prescriptions used in numerical relativity simulations of neutron star mergers to determine whether a given fluid element is unbound and, if unbound, how fast that fluid element will move when far away from the post-merger remnant. We also considered different methods to predict the fallback timescale of material. We emphasize the need to properly take into account heating due to r-process nucleosynthesis and cooling due to neutrino emission on timescales typically not simulated by merger codes ($\sim 1\,{\rm s}$), as well as the thermal energy of hot outflows. We also note that to account for r-process nucleosynthesis, it is crucial to understand the assumptions that go into the construction of an equation of state for the composition of the fluid, including the definition of the mass of a baryon.

Existing prescriptions for the asymptotic properties of matter outflows typically ignore energy losses due to neutrino emission, and sometimes neglect heating due to r-process nucleosynthesis. In this manuscript, we provide an easy-to-use prescription to account for these two effects (Eq.~\ref{eq:hC} and Eq.~\ref{eq:Ginf}) [Model C in the figures], as well as a slight modification to the unbound criterion obtained for that model that accounts for the possibility that ejected material does not have enough time to be heated by the r-process to be unbound before it starts falling back onto the black hole (Eq.~\ref{eq:unb}) [Model D in figures].

Applying our results to existing numerical simulations of BHNS and NSNS mergers, we find that both heating and cooling effects as well as the fluid's internal energy are important to take into account in order to properly predict the mass and kinetic energy of matter outflows. The impact of r-process heating/cooling is particularly important for neutron-rich outflows, i.e. for the matter ejected by the tidal disruption of a neutron star in either NSNS or BHNS mergers. On the other hand, the impact of the heating timescale is found to be relatively small. While model D (Eq.~\ref{eq:unb} and Eq.~\ref{eq:Ginf}) is in theory the most accurate model presented here, model C (Eq.~\ref{eq:hC} and Eq.~\ref{eq:Ginf}) provides similar results with a much simpler method. This is no longer true when considering instead the fallback timescale of bound material. In that case, properly accounting for the finite time over which the r-process occur is crucial to obtain reliable predictions.

Finally, we compare our results to 3D smooth particle hydrodynamics simulations of merger outflows. We find that the simple models presented in this manuscript provide good predictions for the total kinetic energy of the unbound matter, given a heating/cooling model, but that 3D hydrodynamics simulations are required to properly capture the geometry of the outflows. Indeed, energy transfer between fluid elements during the expansion of the outflows lead to a wider spread of velocities (and more spherical outflows) in 3D simulations than what would be predicted when blindly applying the simpler models that can be used in merger simulations. While we recommend the use of a more advanced criteria for the unbound material and its velocity (models C or D) in simulations, we thus note that these models are no substitutes for more costly yet more accurate 3D simulations of the outflows.

\begin{acknowledgments}
The authors thank Stephan Rosswog for useful comments on an earlier version of this manuscript. This research was carried out in part during the 2019 Summer School at the Center for Computational Astrophysics, Flatiron Institute. The Flatiron Institute is supported by the Simons Foundation. 
FF acknowledges
support from the DOE through grant DE-SC0020435,
 from NASA through grant 80NSSC18K0565.
FF and TR acknowledge support from the NSF through grant PHY-1806278. 
This research
was supported in part by the U.S. Department of Energy, Office of Science, Office of Nuclear Physics, under contract number DE-AC02-05CH11231 and DESC0017616, by a SciDAC award DE-SC0018297, and by
the Gordon and Betty Moore Foundation through Grant
GBMF5076. DK acknowledges support from the Simons
Foundation Investigator program under award number 622817. 
This collaborative work was
supported in part by the NSF Physics Frontier Center N3AS under cooperative agreement \#2020275.

\end{acknowledgments}

\bibliography{References/References.bib}

\end{document}